\documentstyle[12pt,aaspp]{article}
\begin{document}

\title{THE OPTICAL DEPTH TO GRAVITATIONAL MICROLENSING IN THE DIRECTION OF
THE GALACTIC BULGE}

\author{A. Udalski\altaffilmark{1}, M. Szyma\'nski\altaffilmark{1},
K. Z. Stanek\altaffilmark{2,6}}
\affil{\tt e-mail I: (udalski,msz)@sirius.astrouw.edu.pl,
stanek@astro.princeton.edu}

\author{J. Ka\l u\.zny\altaffilmark{1}, M. Kubiak\altaffilmark{1},
M. Mateo\altaffilmark{3}, W. Krzemi\'nski\altaffilmark{4}}
\affil{\tt e-mail I: (jka,mk)@sirius.astrouw.edu.pl,
mateo@astro.lsa.umich.edu, wojtek@roses.ctio.noao.edu}

\author{and}

\author{B. Paczy\'nski\altaffilmark{2} and R. Venkat\altaffilmark{5}}
\affil{\tt e-mail I: bp@astro.princeton.edu, rxv8@po.cwru.edu}

\altaffiltext{1}{Warsaw University Observatory, Al. Ujazdowskie 4,
00--478 Warszawa, Poland}
\altaffiltext{2}{Princeton University Observatory, Princeton, NJ 08544--1001}
\altaffiltext{3}{Department of Astronomy, University of Michigan, 821 Dennison
Bldg., Ann Arbor, MI 48109--1090}
\altaffiltext{4}{Carnegie Observatories, Las Campanas Observatory, Casilla
601, La Serena, Chile}
\altaffiltext{5}{Case Western Reserve University, 11896 Carlton Rd. 440B,
        Cleveland, OH 44106}
\altaffiltext{6}{On leave from N. Copernicus Astronomical Center,
Bartycka 18, Warszawa 00-716, Poland}

\begin{abstract}

We present the analysis of the first two years of the OGLE search for
gravitational lenses towards the Galactic bulge.  We detected 9
microlensing events in an algorithmic search of $ \sim 10^8 $
measurements of $ \sim 10^6 $ stars. The characteristic time scales are
in the range $ 8.6 < t_0 < 62 $ days, where $ t_0 = R_E / V $.  The
distribution of amplitudes is consistent with theoretical expectation.
The stars seem to be drawn at random from the overall distribution of
the observed bulge stars. We find that the optical depth to microlensing
is larger than  $ ( 3.3 \pm 1.2 ) \times 10^{-6}$, in excess of current
theoretical  estimates.

\end{abstract}


\section{INTRODUCTION}

The Optical Gravitational Lensing Experiment (OGLE) is a long term
project targeted at the determination of the rate and the statistical
properties of gravitational microlensing of the Galactic Bulge stars.
The project was described by Udalski et al.~(1992) and the first
lensing events were reported by Udalski et al.~(1993b, 1994a).  All
observations were done with the 1 meter Swope telescope at the Las
Campanas Observatory, operated by the Carnegie Institution of
Washington. The detector was a single Loral CCD with ${2048\times2048}$
pixels.  We used a  modified version of DoPhot photometric software
(Schechter, Mateo \& Saha 1993) to extract stellar magnitudes from the
CCD frames.

The purpose of this paper is to present the first estimate of the
optical depth to gravitational microlensing of the Galactic Bulge stars.
In the following sections we present the description of the selection
criteria for microlensing event candidates, the scaling of the
photometric errors as provided by the DoPhot, the analysis of the
efficiency of recovering theoretical microlensing events, the event
statistics, and the overall discussion.  This is largely a technical
paper.  The discussion of our results in the broader astronomical
context will be provided elsewhere (Paczy\'nski et al. 1994b).

The 13 fields covered with our CCD search in the 1992 and 1993
observing seasons are  shown in Fig.~1.  The complete list of
microlensing event candidates is given in Table 1 in the chronological
order of their maximum light $ t_{max}$, which is not the same as the
order in which they were discovered in the data.  The discovery order is
given by the OGLE \#, as listed.

\section{THE LENS CANDIDATES}

The first six OGLE microlensing events as reported by Udalski et al.
(1993b, 1994a) were selected as follows.  The observations were done in
two observing seasons, 1992 and 1993, separated by 8 months during which
no data was taken.  First, the stars constant in season A (be it 1992 or
 1993) were selected from the database of all I-band photometric
measurements with additional requirement that $ I \leq 19.5 $
(Szyma\'nski \& Udalski 1993). The procedure of defining a constant star
is described in detail by Udalski et al.~(1993a, p. 71-73; 1993b). In
short, such a star must have at least $N=12$ ``good'' measurements in
the $I$ band and a standard deviation from the mean less then the
maximum allowed value $\sigma_{max}$ for a given magnitude and a given
field. There were $\sim 1.1 \times 10^6$ such stars in 1992 and $ \sim
1.4 \times 10^6 $ such stars in 1993.  Next, the same stars were looked
at in season B and those which had at least 5 measurements deviating
from the A season average by more than $3\sigma_{max}$ were selected.
Some additional filters were applied to reduce the number of objects.
The remaining ``variable'' stars were all inspected and the six
published events were selected according to a human judgement, not by a
well defined algorithm.  The judgement was relatively simple, as the
vast majority of ``variables'' had very erratic, almost random changes
in magnitude, most likely caused by some defects in the images.  A few
were long period variables.  Only very ``good looking'' events were
finally reported, those clearly standing above the noise.

In order to make a proper estimate of the optical depth to microlensing
we have to know the efficiency of the OGLE system for the detection of
microlensing events of various timescales.  This requires a strictly
algorithmic selection process to be used.  Therefore, we changed a
little the original selection criteria and supplemented them with a
number of  new ones, so as to reduce as much as possible the noise and
to obtain the final, small number of lensing event candidates. Here is
the final list of selection criteria  used.

For a star to be included in the search we required at least $N=15$
good photometric measurements both in season A and B, and at least 5
consecutive measurements in season B that deviated {\it up} in
brightness  by more than $3\sigma_{max}$, or at least 10 such points
total. In addition, if the {\it total} number of such deviant points was
$N_t$ we required that at least $N_c \geq 0.5 N_t$ such measurements be
{\it consecutive}, i.e. we required at least $ N_c $ points to deviate
{\it up}, with no ``low'' points in between. All stars that satisfied
the preceding criteria had their data in season B blindly fitted to a
microlensing light curve of a point mass lens, the magnitude at minimum
light adopted as the mean magnitude in season A.  There were three
adjustable parameters of the model: the peak magnification $A$, the time
of maximum light $t_{max}$, and the characteristic time scale of the
lensing event $ t_0 = R_E / V $, where $ R_E $ is the Einstein ring
radius and $ V $ is the relative transverse velocity in the source --
lens -- observer system (cf. Paczy\'nski 1986).  Next, we required that
there should be at least two data points within $3t_0$ on each side of
$t_{max}$ -- this eliminated a number of ``promising'' light curves near
the beginning or the end of each season, where the fitted maximum fell
beyond the range covered by the observations. Then, the $\chi^2$ was
calculated twice, first with respect to the constant light level given
by season A, and next with respect to the ``best fit'' lensing curve;
the ratio of the  two was required to be in excess of 20.

The distribution of 2041 ``variable'' stars selected according to the
original criteria is  plotted in Fig.~2 in the $ N_t - N_c $ diagram,
together with a line separating them according to $ N_c \geq 0.5 N_t $.
The condition $ N_c \geq 0.5 N_t $ reduced the sample to 469 stars.
Next, we required that at least two points should be located on either
side of the maximum - this reduced the sample to 336 stars.  For these
336 objects the ratio of $ \chi ^2 $ given by the best fit lensing curve
divided by $\chi^2$ calculated with respect to a constant light, versus
$ \chi^2 $ of the fitted lens divided by the number of measurements $N$
is shown in Fig.~3, together with the line separating the final 18
lensing candidates selected algorithmically.

The final step has been done by inspection of the original CCD frames to
check if a particular lens candidate is not a result of some obvious
defect, like a bleeding column.  This has reduced the number of
candidate events to 9, some of them redundant (OGLE \#2a and OGLE \#2b),
as they were measured independently in the regions of overlap of
adjacent CCD fields. All six lenses reported by Udalski et al. (1993b,
1994a) were recovered, as well as the double lens candidate (OGLE \#7,
Udalski et al. 1994b).  Two new events, OGLE \#9 and \#10 have been
found  through the algorithmic search procedure\footnotemark[1]. These
events, in particular \#10, are not nearly as ``good looking'' as those
reported in the past (Udalski et al. 1993b, 1994a).  However, they came
out of the algorithmic search, their CCD images were fine, and there was
no formal reason to reject them. One should be, however, aware that OGLE
\#10 might be in fact a very long period variable star.

The light curves for all final lens candidate events are shown in
Fig.~4.  It is important to note that the photometry we are using
throughout this analysis, and in particular the one shown in Fig.~4 is
the photometry directly available from the database, while the light
curves of the OGLE events published by Udalski et al. (1994a) were based
on refined, differential photometry, as described in that paper.  We use
the database photometry as this is the only one available for all the
stars on which the variety of tests are performed, and we need a uniform
approach to all stars in order to make an unbiased estimate of the
optical depth.

All the events are listed  in Table 1  which gives the field name, the
star number in the database, the total number of measurements $ N_t $
that  deviated up from the constant season average by more than $ 3
\sigma_{max}$ threshold, the number $ N_c $, of such measurements that
were consecutive, the  photometric colors, the time of maximum, the
event time scale $ t_0 $, and other parameters to be described in the
subsequent sections. The last column lists the name of the event.  Seven
events were detected in the 1992 data, while the remaining two events
were detected in the 1993 data.  Notice that the number of stars
searched for lensing was $ \sim 1.4 \times 10^6 $ in 1992 and $ \sim 1.1
\times 10^6 $ in 1993.

\footnotetext[1]{The OGLE lens numbering system
is chronological. The event OGLE \#8 has been discovered
in another data set which is not analyzed in this paper.}

\newpage

\section{SCALING DOPHOT ERRORS}

The limiting magnitude of the OGLE photometry was always determined by
the overlapping star images, never by the photon statistics, and it was
close to the turn-off point of the Bulge main sequence (Udalski et al.
1993a). Therefore, the observed luminosity function was rapidly
increasing with the stellar magnitude,  with most of the stars close to
the detection limit  and even more stars just below. It all implies that
the photometric errors were not easy to estimate, and many stars were
unresolved blends of two or more stellar images. Naturally, we expect
that the majority of the lensing candidates are to be found near the
detection limit and we expect that the measurement errors at their
minimum light are very difficult to estimate. The situation improves as
the stellar image brightens during the event and dominates the
background of faint and unresolved stars.  In the following paragraphs
we describe  our attempt to estimate the errors of our photometry.
However, it should be realized that there is only one way to obtain
truly  reliable photometry for the majority of our lensed stars at their
minimum light: it is necessary to obtain images with a much smaller
seeing, or in other words to move the detection limit by a few
magnitudes below ours.  At this time only HST can provide the required
resolution.

It is important to find the relation between the errors as given by
DoPhot and the real observational errors. To address this issue we have
randomly chosen 1\% of stars (in each of our 13 CCD fields) which were
constant in the first or the second season: $ \sim 11,000 $ in 1992 and
$ \sim 14,000 $ in 1993 and we used them for a variety of tests.  We
shall refer to these as the template stars. Next, we selected those
template stars which had a total of at least 40 photometric measurements
made in the two observing seasons  -- there were $\sim 25,000$ such
stars and we used the measurements from both  seasons for all those
stars in the following analysis.

For every star we calculated the DoPhot error-weighted average
I-magnitude according to: \begin{equation} \bar{I}=\sum_{i}
\frac{I_i}{\sigma_{i,D}}/ \sum_{i} \frac{1} {\sigma_{i,D}},
\end{equation} where $\sigma_{i,D}$ is the DoPhot error of the
measurement number $ i $. The individual deviations were calculated as $
(I_i-\bar{I})/ \sigma_{D}$.  These deviations were grouped according to
the values of $\sigma_{D}$ and $\bar{I}$.  The cells with the number of
deviations exceeding 2,000 were used to obtain scaling factor between
the DoPhot and the real error as a function of $\sigma_{D}$ and
$\bar{I}$. The procedure is described in detail by Lupton et al. (1989),
p. 206, so here we give only its brief outline. From each cell with more
than 2,000 measurements we have randomly selected $ m = 2,000 $
measurements for the analysis so as to have a uniform statistics.  The
2,000 deviations within a given cell were sorted according to the
distance from the mean within the cell and the dispersion of the
deviations was calculated. Next, we removed the deviations farthest from
the mean and we recomputed the dispersion. The whole procedure was
repeated to obtain $s(m,n)$, the dispersion of deviations as a function
of $n$, the number of stars remaining.  Next, the same process was
repeated for deviations drawn randomly from a gaussian normal
distribution and obtained by a series of Monte Carlo simulations.

An example of $ s(m,n)$ for both real and simulated data is shown in
Fig. 5. In the logarithmic plot the curves representing the two
dispersions are parallel up to a certain value of $n$, beyond which the
real data sample is contaminated by variable stars as well as a variety
of defects in the CCD frames. This means the DoPhot error  $\sigma_{D}$
may represent the real observational error provided it is multiplied by
an appropriate scaling factor F  which in the case shown in Fig. 5 is $
F = 1.48 $, and corresponds to the average distance between the two
parallel curves. The same procedure was used to obtain the scaling
factor for all $ (\sigma_{D} , \bar{I})$ cells with more than 2,000
deviations. For the cells with fewer than 2,000 deviations the average
scaling factor $ F = 1.29 $ was adopted. The values of F-factors are
listed in Table 2 as a function of $ \bar I $ and $ \sigma_D $ for each
cell which had more than 2,000 measurements in it.  The number of
measurements per cell is given in Table 3.

Fig. 5 can also be used to estimate the fraction of measurements that
correspond to either variable or ``defective'' star images. The $
s(m,n)$ curve for the real  data deviates from the simulated one at
about $ n=1,800 $, which indicates that roughly 10\% of all measurement
are non-gaussian for whatever reason. It seems likely that the vast
majority of these ``variables'' are spurious, indicative of severe
crowding and a variety of CCD defects. It is worth noticing that the
same analysis when performed for all stars in the database of
Szyma\'nski \& Udalski (1993) revealed $ \sim 30\% $ of non-gaussian
measurements.  This indicates that the  procedure of pre-selecting
constant stars removed most of the noise from  the data.  Unfortunately,
even the $ \sim 10\% $ contamination of the  measurements of constant
stars by the non-gaussian tail does not allow us to use rigorous $ \chi
^2 $ tests to assess the quality of agreement between the candidate
lensing events and the best fit theoretical curves.

As a byproduct of finding the scaling factors we also found a relation
between the DoPhot errors and the stellar magnitude.  For each bin of
the I magnitude we found the most common value of $ \sigma_D $ in Table 3
and we  fitted the relation with the simple formula:
\begin{equation}
\sigma_{D,I_1} \approx \sigma_{D,I_2}10^{(I_1-I_2)/3.5},
\end{equation}
which allows us to estimate how the measurement error scales with
stellar brightness.  This is important in the models of lensing events.
According to eq. (2) when a star brightens from $ I_2 $ to $ I_1 $ by
3.5 magnitudes then its DoPhot error decreases by a factor $\sim 10 $.
This relation will be used in our Monte Carlo estimate of the OGLE
microlensing detection efficiency.

\section{THE EFFICIENCY OF RECOVERY OF MICROLENSING EVENTS}

The efficiency of the algorithmic search procedure described in section
2 was tested with $ \sim 14,000 $ stars in 1992, and with $ \sim 11,000
$ stars in 1993. The template stars were not necessarily constant in the
test season  as they were selected as constant in the other season. They
provided the time sequences of measurements and their DoPhot errors.
For every value of event time scale $t_0$ we generated 100,000 Monte
Carlo simulations of microlensing events. The star to be microlensed was
randomly selected each time and the dimensionless impact parameter $
p/R_E $ and the time of peak magnification $ t_{max}$ were randomly
selected from the uniform distributions: $ 0 \leq p/R_E \leq 2 $, $ 0
\leq t_{max} \leq 1 ~ yr $. $ R_E $ is the Einstein ring radius, and $
t_{max} = 0 $ corresponded to either 1992.0 or 1993.0, for the two
observing seasons, respectively. The simulations were done for 61 values
of $ t_0 $ in the range  $ -1 \leq \log t_0 \leq 2 $ (days).

The data points for simulated events were obtained as follows. First,
theoretical magnitude $ I_t $ was calculated from a model  for every
value of time for which a real measurement and its DoPhot error were
available, $ I_{obs}$, $ \sigma_{D,obs}$.   Next, the DoPhot error $
\sigma_{D,t}$ was assigned to the theoretical  magnitude by rescaling
the observational error $ \sigma_{D,obs}$ according to the magnitude
difference $ I_t - I_{obs}$ and following eq. (2).  The ``true''
theoretical error was calculated as $ \sigma_t = \sigma_{D,t} \times F
$, where the scaling factor was obtained from Table 2 according to the
values of $ I_t $ and $ \sigma_{D,t}$.  Finally, the ``actual''
theoretical error was obtained with the Monte Carlo simulation using
gaussian distribution with the standard deviation given as $ \sigma_t $
and this ``actual'' error was added to the theoretical magnitude $ I_t
$ to obtain the simulated data point $ I_s $.  This procedure was
repeated to generate a series of simulated data points for every time at
which there was an actual measurement available.  The full series of
simulated data points was treated in the same way as a series of real
observations and the algorithm described in section 2 was used for
detection of the model event.  Four examples of simulated events are
shown in Fig. 6.

The fraction of events that have passed the detection criteria was
multiplied by a factor two  and adopted as the OGLE efficiency to
detect events of a given time scale $t_0$.  This factor  comes from the
fact that the traditional definition of the optical depth requires the
source to be within one Einstein radius of the lens, i.e. only the
events with the dimensionless impact parameters smaller than unity,
$p/R_E \leq 1 $ are counted, while our simulations covered a region twice
as large, $ p/R_E \leq 2 $.  Imagine, that we were able to recover {\it
all} events with $ p/R_E \leq 1 $ and none with a larger impact
parameter. In this case 50\% of all simulated events would be detected
but the OGLE efficiency would be 100\%.

The OGLE detection criterion is expressed in terms of measurement errors
rather than the event magnification.  A low amplitude event may be
detected if the source is bright and the errors are small, like the
OGLE~\#3  for which $ p/R_E = 1.08 $, whereas a microlensing of a very
faint star may be detectable only for $ p/R_E \ll 1 $. If the OGLE
experiment had no gaps in the data, the observations were carried out
for 12 months every year, and we were able to recover all events with $
p/R_E \leq 2 $, then our sensitivity would be equal 2.  Of course, in
that case we would have to extend our simulations to larger impact
parameters, until we reach so low magnifications that the model events
are  no longer  recoverable.

The OGLE observing season lasted about 4 months every year with large
gaps in the coverage caused by the telescope scheduling and the weather
conditions.  Hence, the sensitivity was never even close to 1, not to
mention 2.  The plots of the efficiency as a function of event timescale
is shown in Fig. 7 for the 1992 and 1993 seasons with the solid and
dashed lines, respectively.  The two lines are rather close to each
other.  The efficiency drops rapidly for $ t_0 $ shorter than a few days
as the stars were never observed more than twice a night.  The
efficiency approaches $ \sim 30\% $ for long timescales, which is
approximately the fraction of each year covered by the OGLE observing
season.  Please note that at present we  do not investigate the
efficiency of recovery of the very long lasting events with $t_0>1\;yr$,
as those would be classified as variable stars in both 1992 and 1993.
We plan to investigate those long  timescales when the data covering
many years becomes available.

\section{THE EVENT STATISTICS}

Because of distinctly non-gaussian distribution of the OGLE measurement
errors  even for the stars selected as constant, we cannot use the
standard $ \chi ^2 $ test to assess the goodness of fit between the
observed luminosity variations of the microlensing candidate events.
Therefore, we have to make other checks of the statistical properties we
expect of genuine lensing events.  The following test is equivalent to a
comparison between the observed distribution of event amplitudes  and
the distribution expected theoretically.  The next two tests are
designed to verify the expectation that all Galactic Bulge stars  are
equally likely to be microlensed.

\subsection{ $ p/p_{max} = u/u_{max}$ statistics}

A question frequently asked is: do OGLE events have the expected
distribution of peak magnifications?  This cannot be answered directly
as the detection threshold is very fuzzy.  Instead, we may use a
criterion similar to the $ \langle V/V_{max} \rangle $ originally
proposed by Schmidt~(1968) to study quasar distribution.

For every candidate event the best fit provides a value of the
dimensionless impact parameter, $ u \equiv p/R_E $, and that is directly
related to the peak magnification $A$:
\begin{equation}
A = { u^2 + 2 \over u \sqrt{ u^2 +4 } } ~~ ,
\end{equation}
(e.g. Paczy\'nski 1986).  Given the particular lens case with its
distribution of measurements and errors we may ask a question: what is
the maximum impact parameter $ u_{max}$ for which this event would have
passed our algorithmic detection criteria as described in section 2? Of
course, the larger the impact parameter the lower the magnification and
the closer are all measurements to the baseline magnitude.

Following the description given in section 4 a series of simulated
events was generated with all lens parameters kept constant, except for
the dimensionless impact parameter $u$,  The procedure was repeated
until the maximum value $ u_{max}$ was found; this corresponded to the
detection threshold of the event. As the test events were Monte Carlo
simulated there was not a discontinuous jump from full detectability for
$ u \leq u_{max}$ to none above $ u_{max}$.  Rather, the fraction of
model events that were detected varied fairly rapidly but smoothly with
$ u $. Therefore, the $ u_{max}$ was defined with the integral formula
\begin{equation}
u_{max} = \int _0^{\infty} f_d (u) ~ du  ,
\end{equation}
where $ f_d (u)$ is the fraction of events detected as a function of the
impact parameter.  Naturally, in practice there was never any need to
extend the integral to infinity, as the detectability $ f_d (u)$
approached zero very rapidly as soon as the impact parameter exceeded $
u_{max}$.

The lens trajectory with respect to the source is expected to be random.
Thus, a detectable lens has its impact parameter uniformly distributed
in the interval $ 0 \leq u \leq u_{max}$, and therefore we expect that
$u/u_{max}$  should be uniformly distributed in the range (0,1).
Notice, that the specific value of $ u_{max}$ has to be determined for
each event individually, as every event has a unique time sequence of
measurements and errors.  Nevertheless, we expect all events to share
the property that their $ u/u_{max}$ parameters should be uniformly but
randomly distributed in the (0,1) interval. Indeed, we find that $
u/u_{max} $ is uniformly distributed as expected for random impact
parameters - this point will be discussed in subsection (5.4)

The situation is somewhat more complicated if a lens is double. In this
case a concept of positive and negative impact parameter has to be
introduced, and the actual impact parameter is expected to be uniformly
distributed in the range
$ u^- \leq u \leq u^+ $, where $ u^- $ and $ u^+ $ are the minimum and
the maximum values for which the lens can be detected.  One of our
events, OGLE \#7, has a very dramatic light curve and it is likely to be
caused by a double lens (Udalski et al. 1994b).  In this particular case
the absolute values of $ u^- $ and $ u^+ $ are about equal, and it makes
sense to use the $ u/u_{max}$ parameter. Another possibly double lens,
OGLE \#6 (Mao \& DiStefano 1994), is not so dramatic and for the purpose
of the current analysis we treat it as a single lens case.

The best fit dimensionless impact parameters, $u$, and the maximum
dimensionless impact parameters $u_{max}$ are listed  in Table 1 for all
OGLE events. For the purpose of this analysis we treated OGLE \#7 as a
single lens, thus all its parameters given in Table 1 should be treated
as crude estimates only.  A full double lens analysis will be presented
elsewhere (Udalski et al. 1994b).

\subsection{Detectability distribution}

Now we have to check if the lensed stars are random  or they have some
common properties which might indicate that the variability has
something to do with the star.  Gravitational lensing should affect all
Galactic Bulge stars with the same probability.

In this subsection we check if the lens candidates are randomly
distributed in terms of their ``detectability''.  Let us take a specific
OGLE event \#k, where k = 1, 2, 3, 4, 5, 6, 7, 9, 10, with its specific
timescale $ t_{0,k}$ and its dimensionless impact parameter $u_k$.  For
every event we repeated our sensitivity test as described in section 4
for the specific values of $t_{0,k}$ and $u_k$.  We made 200 simulations
for each of the $\sim 14,000 $ template stars which  were selected as
constant in 1993 to test them for the ``detectability'' of OGLE lenses
\# 2, 3, 4, 5, 6, 9 and 10 in the 1992 data.  We also made 200
simulations for each of the $\sim 11,000 $ template stars which  were
selected as constant in 1992 to test them for the ``detectability'' of
OGLE lenses \#1 and 7 in the 1993 data.

In addition we made $4,000$ simulations  for each star which was
observed as the candidate event OGLE \#k with its time scale $ t_{0,k}$.
 For every star, including the  one that had undergone the event, we
calculate the fraction $ \epsilon_i = N_{r,i}/N_{t,i}$.  $N_{t,i}$ is
the total number of  simulated events for the star number $i$; $N_t=200
$ for test stars  and $ N_t=4,000$ for the event star.  $ N_{r,i}$ is
the number of cases in which a theoretical event has been recovered for
the star number $i$. We followed the procedure described in section 2 to
``detect'' these theoretical events.  The larger the value of $
\epsilon_i $ the easier  it was to recover the events, i.e. the
``better'' was the star.

Now, we rank order stars according to the value of their  $ \epsilon_i $
parameter, and we establish a cumulative  recovery probability  as a
function of $ \epsilon $, the efficiency of detection:
\begin{equation}
P(\leq \epsilon) =
\sum_{i, (\epsilon_i \leq \epsilon) } \epsilon _i / \sum_{i} \epsilon _i ~~~ ,
\end{equation}
$P(\leq \epsilon)$ is the probability that an event with the timescale
$t_{0,k}$ and the dimensionless impact parameter $ u_k $ is discovered
among the stars for which the discovery efficiency  is less or equal $
\epsilon $.  Finally, we calculated $P(\leq \epsilon_k)$, where $
\epsilon_k $ was the recovery fraction for the actual OGLE \#k event,
the one we were analyzing. If its selection was fair then the value of
$P(\leq \epsilon_k)$ should have a uniform probability distribution in
the interval (0,1). The values of $P(\leq \epsilon_k)$ are listed in
Table 1. The relations between $P(\leq \epsilon)$ and $ \epsilon $ are
shown for all OGLE events in the nine panels of Fig. 8. The vertical
dashed lines correspond to the the value of $ \epsilon_k $  for each
lens.

\subsection{I magnitude distribution}

We followed a similar procedure to find out if the candidate events were
randomly but uniformly distributed in  the I magnitude of our sample.
 For every template star we had the value of its magnitude $ I_i $ and
its efficiency parameter for the lens detectability $ \epsilon_i $. The
value of $ \epsilon _i $ was specific to every OGLE event, as described
in previous subsection.  We rank ordered all stars  according to their
$ I_i $ magnitude and we calculated

\begin{equation}
P(\leq I) = \sum_{i, ( I_i \leq I) } \epsilon_i / \sum_i \epsilon_i ,
\end{equation}

where $ P(\leq I)$ is the probability that an event with the time scale
$ t_{0,k}$ is discovered among stars brighter than $I$.  The value
corresponding to the event OGLE \#k is  $ P(\leq I_k)$, where $I_k$ is
its baseline magnitude. If the selection was fair then the value of
$P(\leq I_k)$ should be uniformly but randomly distributed in the
interval (0,1). The values of $P(\leq  I_k)$ are listed in Table 1. The
relations between $P(\leq  I)$ and $  I $ are shown for all OGLE events
in the nine panels of Fig. 9. The vertical dashed lines correspond to
the the value of $  I_k $ for each lens.

The efficiency of lens detectability $ \epsilon $  is in general higher
for bright stars because their measurements have smaller errors, so we
expect that the last two distribution tests are somewhat related.  The
correlations between the values of $ \epsilon $ and I magnitude is shown
in nine panels in Fig. 10 for the nine lens candidate events.  Notice,
that while in many cases the correlation is strong it is either weak or
absent in some cases.  In general, it is clear that each lens has a
different detectability pattern among the stars.

\subsection{Overall distribution properties}

The results of the three preceding tests are conveniently displayed in
three panels in Fig. 11.  The distribution of all three parameters
should be uniform and random in the range (0,1) and it seems to be such.
With the small number of events there is no point to apply a
sophisticated statistical analysis -- it is clear that there are no
significant departures from the expected distribution.  In other words,
we have no reason to doubt that the 9 events presented in this paper are
not due to gravitational microlensing.

Still, one may notice that there are no events in the middle panel of
Fig. 11 for  $ 0.75 \leq P(\leq \epsilon) \leq 1.0 $, and none in the
lower panel for $ 0.0 \leq P( \leq I ) \leq 0.17 $.  These two are
correlated as explained in the previous sub-section: bright stars have
higher efficiency for lens detection. The gaps are not statistically
significant, but they may be partly due to a relatively large
contribution of the galactic disk at the bright end.  Disk stars make
$ \sim 50\% $ of all stars brighter than $ V = 15 $, and $ \sim 20\% $ of
all starsbrighter than $ V = 18 $.  We have not excluded them from the
search.  We did not expect any lensing events among them, and we found
none.  This absence may contribute to the apparent gaps in the
distribution in the middle and the lower panels of Fig. 11.

\subsection{Color distribution}

We cannot easily test the distribution in $ V-I $ colors, as the
database of stars for which good $ V-I $ color are known is about three
times smaller than the I magnitude database. The reason is simple: most
stars are close to the detection limit, and due to large reddening of
the Galactic bulge  many of them have only I magnitude well measured,
while only some have both. Nevertheless, a qualitative impression of a
random distribution is apparent when the locations of the OGLE lensed
stars are plotted  in the color -- magnitude diagram, as shown in Fig.
12. The positions of $ \sim 12,000 $ stars are plotted, 2.5\% of the
whole database for which we have good color information in our 13
fields. The final lensing candidates are shown with large circles; their
distribution seems to be random, and clearly belonging to the bulge
population.

\subsection{Color variations}

The OGLE experiment has only limited color information as most
measurements were made in the I-band, and only occasionally in the
V-band (Udalski et al. 1992, 1994a).  No color change was detected for
any of the events, but it should be kept in mind that for many stars the
colors at minimum light are uncertain, as those stars are close to the
detection limit.  However, the OGLE \#3 event was so bright that the
color was found to be constant to better than 0.02 magnitude.

\section{THE OPTICAL DEPTH TO MICROLENSING}

Let us suppose that all microlensing events have the same time scale
$ t_0 $, and that the detection efficiency is 100\%.  The frequency
of events per year $ \Gamma $ is related to the average time between the
events $ \langle \Delta t \rangle $ as

\begin{equation}
\Gamma = { 1 ~ yr \over \langle \Delta t \rangle } ,
\end{equation}

and the optical depth can be calculated as (cf. Paczy\'nski 1986)

\begin{equation}
\tau = \left( { \pi \over 2 N_m } \right)
\left( { t_0 \over \langle \Delta t \rangle } \right)  ,
\end{equation}

where $ N_m $ is the number of stars that are monitored, and we count
only those events for which the impact parameter is smaller than the
Einstein ring radius.  In fact the OGLE efficiency is much less than
100\% and for each event we have a different time scale $ t_0 $ and a
different maximum impact parameter to which that event would be
detectable.  We consider the whole set of events covering the two
observing seasons.  Therefore, the contribution of event \#k to the
overall optical depth is divided by a factor two (2 years),

\begin{equation}
\tau _k = \left({ \pi \over 2 N_m } \right)
\left({ t_0 \over \epsilon \times 2 ~ yr } \right)
\left({ 1 \over u_{max} } \right) ,
\end{equation}

where $ \epsilon $ is the OGLE efficiency corresponding to the event
time scale $ t_0 $, as shown in Fig. 7, $ N_m $ is the number of
``constant'' stars that were effectively monitored in the particular
season, and $ u_{max} = p_{max}/R_E $.

The total number of stars in the database of constant stars with $ I
\leq 19.5 $ was $\sim 1.4 \times 10^6 $ in 1992 and $ \sim 1.1 \times
10^6 $ in 1993. However, these are not independent for technical
reasons. First, as a consequence of PSF varying over the
${2048\times2048}$  CCD chip it was necessary to divide each frame into
49 sub-frames, as described by Udalski et al. (1992).  This made some
stars listed twice (sometimes more than twice) in the database, and  all
measurements were made separately for each entry.  The cross-link is
available,  and the true number of stars has to be reduced by $ \sim
12\% $ to account for the multiple listings.  We have conducted the
search on all listed objects, and in this way the OGLE \#1 has been
detected twice, under two names of the same star in the same Baade's
Window field BW7.

Next, there was some degree of overlap between adjacent CCD fields,
as described by Udalski et al. (1992).  This made $ \sim 12\% $
of all stars measured twice as they appeared in the overlap regions.
This way the OGLE \#2 has been detected in Baade's Window fields
BW5 and BWC (cf. Table 1).

Finally, some stars were not in the Galactic bulge but in the galactic
disk, on average too close to be lensed.  We very crudely estimate their
contribution to be $ \sim 5\% $, though it may be somewhat larger, as
most disk stars are presumably blended with the bulge stars in the
region of the color magnitude diagram where most of all stars are
located: $ V \approx 20 $, $ V-I \approx 1.6 $, as shown in Fig. 12.

All three effects combined act in the same direction: they reduce the
effective number of stars searched for microlensing by a factor
$ \sim 1.47 $.  Therefore, the numbers to be used in evaluating
the optical depth with the eq. (9) should be
$ N_m \approx 0.95 \times 10^6 $ in 1992 and
$ N_m \approx 0.75 \times 10^6 $ in 1993.
The contributions of the final candidate events to the optical depth (in
units of $ 10^{-6}$) are given in Table 1.  The combined optical depth
is $ \tau \approx ( 3.3 \pm 1.2 ) \times 10^{-6}$, where
the standard deviation was calculated according to the formula:
\begin{equation}
\sigma_{ \tau } = \left( \sum_k ( \tau _k )^2 \right) ^{1/2} ~ .
\end{equation}
Of course, this is only a random error calculated as if all events
were independent.  It does not allow for any systematic
errors.  We tried to minimize their effect but it is difficult
to assess at this time how successful we were in this task.

\section{DISCUSSION}

Our estimate of the optical depth to gravitational microlensing towards
the galactic bulge was made treating all 13 OGLE fields as equal.  This
includes 9 Baade's Window fields at $ l = 1^o $, $ b = -4^o $, and 4
Galactic Bar fields at about the same galactic latitude but at $ l = \pm
5^o $ (cf. Fig. 1). If the microlensing is dominated by the galactic
bulge lenses as asserted by Kiraga and Paczy\'nski (1994) then the
expected rate should be less in Galactic Bar fields than in BW fields.
Our results are consistent with this notion, but they are just as
consistent with the rate being the same in all fields (cf. Table 1) as
the number of events detected is small.

In order to check how our estimate of the optical depth depends on our
ad hoc choice of the detection threshold as described in section 3, we
increased the threshold from $ 3 \sigma_{max} $ to $ 8 \sigma _{max} $,
keeping all other rules of the algorithm the same. As expected many
lenses dropped out, and only 4 remained: \#3, \#2a, \#5, and \#7.  The
new efficiencies $ \epsilon _k $ and the new values of the maximum
impact parameters $ u_{max} $ were recalculated as well, and the new
estimate of the optical depth was obtained: $ \tau = ( 6.8 \pm 4.4)
\times 10^{-6} $.  This is within one standard deviation of the estimate
based on the $ 3 \sigma _{max} $ detection threshold, indicating that
the result is not sensitive to the choice of the threshold.

Our estimate of the optical depth to gravitational microlensing towards
the galactic bulge is only a lower limit.  The OGLE is not sensitive to
events with the time scale much less than 10 days, while in a recent
theoretical model by Kiraga and Paczy\'nski (1994) the rate of
microlensing events is likely to peak at $ t_0 \sim 10 $ days, even if
there are no brown dwarfs in the Galaxy.  Also, the OGLE is not
sensitive to events lasting longer than the observing seasons, with $
t_0 \geq 100 $ days.  Such events were discriminated against with our
search procedure.  This would not miss many ordinary stellar mass
lenses, but a significant population of dark objects with masses well
above solar would be missed.

It should be pointed out that the rate of events detected by the OGLE,
even though it is only a lower limit, is well in excess of any
theoretical prediction todate -- those were roughly in the range  $ 0.5
\times 10^{-6} \leq \tau \leq 1.0 \times 10^{-6}$ (Paczy\'nski 1991,
Griest et al. 1991, Kiraga and Paczy\'nski 1994).  The discussion of the
astronomical consequences of our finding will be published elsewhere
(Paczy\'nski et al. 1994b).

It should also be noted  that many of the stars which were measured
as single are in fact unresolved blends of a few stars. Notice, that the
seeing disk is typically $ \sim 1'' $, while the cross-section for
gravitational microlensing is typically $ \sim (0.001'')^2 $. This
leaves plenty of room for unresolved blends of which only one  is to be
lensed.  If that happens the amplitude as measured is reduced and the
impact parameter for the lensing is overestimated. In addition, our
experience with improved photometry (Udalski et al. 1994a) indicates
that with the corrected stellar position the amplitudes turns out to be
larger, i.e. in the original database photometry the amplitudes are
underestimated, and the impact parameters are overestimated. Both
effects act in the same direction and lead to underestimate of the
actual number of lensing events and to underestimate of the optical
depth to microlensing.  We have made no quantitative assessment of this
effect, but it can only increase even more the apparent discrepancy
between the observation and currently available models.

Photometry of the OGLE microlensing events, as well as a regularly
updated OGLE status report can be found over the Internet from
``sirius.astrouw.edu.pl'' host (148.81.8.1), using the ``anonymous ftp''
service (directory ``/ogle'', files ``README'', ``ogle.status''). The
report contains the latest news and references to all OGLE related
papers, and the PostScript files of some publications, including
this one. Information on the recent OGLE status is also available via
"World Wide Web" WWW: "http://www.astrouw.edu.pl/".

\acknowledgments{This project was supported with the NSF grants AST
9216494 and AST 9216830 to B. Paczy\'ski and Polish KBN grants No
2-1173-9101 and BST475/A/94 to M. Kubiak.}

\begin{figure}
\begin{center}
{\bf FIGURE CAPTIONS}
\end{center}

\caption{Positions in the Galactic coordinates of 13 fields in which
a search for gravitational microlensing was carried out with the OGLE
in 1992 and 1993 observing seasons.}

\caption{The distribution of stars that were constant in one season and
variable in the other season in the $ N_t - N_c $ plane.  The bigger the
symbol the larger number of stars are at that location. $ N_t $  is the
total number of measurements deviating {\it up} from the constant season
average by more than 3 standard deviations; $ N_c $  is the maximum
number of such measurements that are consecutive, i.e. with no
non-deviating points between them.  The dashed line corresponds to $ N_c
= 0.5 N_t $. Stars below this line were excluded from further analysis.}

\caption{The distribution of stars constant in one season and variable
in the other in the $\chi ^2 $ plane. $\chi^2 $ is the sum of squares of
deviations from the best fit microlensing curve in units of the  DoPhot
errors. $\chi^2_{const}$ is the sum of squares of deviations from the
constant season magnitude, also in units of the DoPhot errors. $N$ is
the number of good measurements in the ``variable'' season.  Only the
stars below the horizontal dashed line were selected for farther
analysis.}

\caption{The observed and theoretical (short dashed lines) light curves
are shown for all events that were below the dashed horizontal line in
Fig. 3 for which no obvious defects were found in the CCD frames.
The solid horizontal lines show the level of average I magnitude in the
observing season when the particular star was constant. The two
horizontal dashed lines are separated from the solid lines by $
\sigma_{max}$ as described in section 2.}

\caption{The variation of the variance $s(m,n)$ as a function of the
number of measurements $n$ remaining in the sample of originally
$m=2,000$ measurements. The upper dashed line is based on real OGLE data
with the DoPhot errors and the I-band magnitudes in the ranges $ 0.016 <
\sigma_D \leq 0.024 $ and $ 17.68 < \bar{I} \leq 17.83 $, respectively.
The solid line corresponds to a theoretical relation for a  gaussian
distribution with a standard deviation $\sigma=1.0$.  The error bars
along the solid line correspond to one standard deviation as estimated
from a large number of Monte Carlo simulations. The lower dashed line is
just the upper line shifted down by $0.17$ in the $ \log s $  which
corresponds to the scaling factor $ F = 1.48$, as described in the text.
Notice, that for $ n \geq 1,800 $ the down-shifted dashed line deviates
upwards from the theoretical line. This indicates that there is a $ \sim
10\% $ non-gaussian tail of the OGLE errors.}
\end{figure}

\begin{figure}
\caption{Four examples of template star measurements are shown with open
circles and the four simulated microlensing events are shown with filled
circles and short dashed lines.
}

\caption{The sensitivity of the OGLE microlensing search $ \epsilon $ is
shown as a function of the event timescale $ t_0 $ for the two observing
seasons.  Notice the abrupt drop in sensitivity for timescale below few
days, and the leveling off at $\sim 30\% $  for long timescale events.}

\caption{The distributions of cumulative probability $ P(\leq \epsilon)$
for the detection of a lensing event of a given time scale $ t_0 $ is
shown as a function of the detection efficiency $ \epsilon $ for   $ \sim
14,000 $ template stars in 1992 and $ \sim 11,000 $ template stars
in 1993. Each panel corresponds to a different lensing event and its
time scale $ t_0 $. The efficiency of detection for each event is
indicated with a vertical dashed line.
}

\caption{ The distributions of cumulative probability $ P(\leq  I)$ for
the detection of a lensing event of a given time scale $ t_0 $ is shown
as a function of I-magnitude for  $ \sim 14,000 $ template stars in
1992 and $ \sim 11,000 $ template stars in 1993.  Each panel corresponds
to a different lensing event and its time scale $ t_0 $.  The
I-magnitude for  each event is indicated with a vertical dashed line.}

\caption{ The distribution of values of the efficiency
parameter $ \epsilon $ and the stellar I-band magnitude is shown.
Notice that while I magnitude is always the same, the efficiency for a
given star depends on which lens is being tested.}

\caption{ The distribution of the statistical parameters for the 9 OGLE
lens candidates: $ u/u_{max}$, $ P(\leq \epsilon)$, and $ P(\leq I)$.
The parameters are explained in sections (5.1-5.3), and their values are
listed in Table 1.  The OGLE lens number is given above each bar
representing the event.  The events should be randomly distributed in
the interval (0,1) in all three panels if the events are due to
gravitational microlensing.}

\caption{The distribution of final lensing candidates (large circles)
is shown in the color -- magnitude diagram together with   $ \sim
12,500 $ stars randomly chosen from the color -- magnitude  diagrams of
all fields. Notice the ``blue main sequence'' of the disk stars (cf.
Paczy\'nski et al. 1994).}
\end{figure}

\end{document}